\documentclass{emulateapj}
\usepackage{apjfonts}
\usepackage{rotating}

\lefthead{CHOI ET AL.} 
\righthead{MICROLENSING BINARY BROWN DWARFS}

\newcommand{\pivec}{\mbox{\boldmath $\pi$}}

\begin{document}
\title{Microlensing Discovery of a Population of Very Tight, Very Low-mass Binary Brown Dwarfs}

\author{
J.-Y. Choi$^{1}$ ,
C. Han$^{1,101,112}$ ,
A. Udalski$^{2,102}$ ,
T. Sumi$^{3,103}$ ,
B. S. Gaudi$^{4,101}$ ,
A. Gould$^{4,101}$ ,
D. P. Bennett$^{5,103}$ ,
M. Dominik$^{6,104,105,111}$ ,
J.-P. Beaulieu$^{7,106}$ ,
Y. Tsapras$^{8,9,105}$ ,
V. Bozza$^{10,11,104}$ , \\
and \\
F. Abe$^{12}$ ,
I. A. Bond$^{13}$ ,
C. S. Botzler$^{14}$ ,
P. Chote$^{15}$ ,
M. Freeman$^{14}$ ,
A. Fukui$^{16}$ ,
K. Furusawa$^{12}$ ,
Y. Itow$^{12}$ ,
C. H. Ling$^{13}$ ,
K. Masuda$^{12}$ ,
Y. Matsubara$^{12}$ ,
N. Miyake$^{12}$ ,
Y. Muraki$^{12}$ ,
K. Ohnishi$^{17}$ ,
N. J. Rattenbury$^{14}$ ,
To. Saito$^{18}$ ,
D. J. Sullivan$^{15}$ ,
K. Suzuki$^{12}$ ,
W. L. Sweatman$^{13}$ ,
D. Suzuki$^{3}$ ,
S. Takino$^{12}$ ,
P. J. Tristram$^{19}$ ,
K. Wada$^{3}$ ,
P. C. M. Yock$^{14}$  \\
 (The MOA Collaboration), \\
M. K. Szyma\'nski$^{2}$ ,
M. Kubiak$^{2}$ ,
G. Pietrzy\'nski$^{2,20}$ ,
I. Soszy\'nski$^{2}$ ,
J. Skowron$^{4}$ ,
S. Koz{\l}owski$^{2}$ ,
R. Poleski$^{2}$ ,
K. Ulaczyk$^{2}$ ,
\L. Wyrzykowski$^{2,21}$ ,
P. Pietrukowicz$^{2}$  \\
(The OGLE Collaboration) \\
L. A. Almeida$^{22}$ ,
D. L. DePoy$^{23}$ ,
Subo Dong$^{24}$ ,
E. Gorbikov$^{25}$ ,
F. Jablonski$^{22}$ ,
C. B. Henderson$^{4}$ ,
K.-H. Hwang$^{1}$ ,
J. Janczak$^{26}$ ,
Y.-K. Jung$^{1}$ ,
S. Kaspi$^{25}$ ,
C.-U. Lee$^{27}$ ,
U. Malamud$^{25}$ ,
D. Maoz$^{25}$ ,
D. McGregor$^{4}$ ,
J. A. Mu$\tilde{\rm n}$oz$^{28}$ ,
B.-G. Park$^{27}$ ,
H. Park$^{1}$ ,
R. W. Pogge$^{4}$ ,
Y. Shvartzvald$^{25}$ ,
I.-G. Shin$^{1}$ ,
J. C. Yee$^{4}$  \\
(The $\mu$FUN Collaboration) \\
K. A. Alsubai$^{29}$ ,
P. Browne$^{6}$ ,
M. J. Burgdorf$^{30}$ ,
S. Calchi Novati$^{31}$ ,
P. Dodds$^{6}$ ,
X.-S. Fang$^{32}$ ,
F. Finet$^{33}$ ,
M. Glitrup$^{34}$ ,
F. Grundahl$^{34}$ ,
S.-H. Gu$^{32}$ ,
S. Hardis$^{35}$ ,
K. Harps{\o}e$^{35,36}$ ,
T. C. Hinse$^{27,35}$ ,
A. Hornstrup$^{37}$ ,
M. Hundertmark$^{6,38}$ ,
J. Jessen-Hansen$^{34,39}$ ,
U. G. J©ªrgensen$^{35,36}$ ,
N. Kains$^{6,40}$ ,
E. Kerins$^{41}$ ,
C. Liebig$^{6,42}$ ,
M. N. Lund$^{34}$ ,
M. Lundkvist$^{34}$ ,
G. Maier$^{42}$ ,
L. Mancini$^{11,43}$ ,
M. Mathiasen$^{35}$ ,
M. T. Penny$^{4,41}$ ,
S. Rahvar$^{44,45}$ ,
D. Ricci$^{33,46}$ ,
G. Scarpetta$^{10,31}$ ,
J. Skottfelt$^{35}$ ,
C. Snodgrass$^{47,48}$ ,
J. Southworth$^{49}$ ,
J. Surdej$^{33}$ ,
J. Tregloan-Reed$^{49}$ ,
J. Wambsganss$^{42}$ ,
O. Wertz$^{33}$ ,
F. Zimmer$^{42}$  \\
(The MiNDSTEp Consortium) \\
M. D. Albrow$^{50}$ ,
E. Bachelet$^{51}$ ,
V. Batista$^{4}$ ,
S. Brillant$^{48}$ ,
A. Cassan$^{52}$ ,
A. A. Cole$^{53}$ ,
C. Coutures$^{52}$ ,
S. Dieters$^{53}$ ,
D. Dominis Prester$^{54}$ ,
J. Donatowicz$^{55}$ ,
P. Fouqu\'e$^{51}$ ,
J. Greenhill$^{53}$ ,
D. Kubas$^{48,52}$ ,
J.-B. Marquette$^{52}$ ,
J. W. Menzies$^{56}$ ,
K. C. Sahu$^{57}$ ,
M. Zub$^{42}$  \\
(The PLANET Collaboration)\\
D. M. Bramich$^{58}$ ,
K. Horne$^{6}$ ,
I. A. Steele$^{59}$ ,
R. A. Street$^{8}$  \\
(The RoboNet Collaboration)
}

\affil{$^{1}$Department of Physics, Institute for Astrophysics, Chungbuk National University, Cheongju 371-763, Republic of Korea} 
\affil{$^{2}$Warsaw University Observatory, Al. Ujazdowskie 4, 00-478 Warszawa, Poland} 
\affil{$^{3}$Department of Earth and Space Science, Osaka University, Osaka 560-0043, Japan} 
\affil{$^{4}$Department of Astronomy, Ohio State University, 140 West 18th Avenue, Columbus, OH 43210, USA} 
\affil{$^{5}$University of Notre Dame, Department of Physics, 225 Nieuwland Science Hall, Notre Dame, IN 46556-5670, USA} 
\affil{$^{6}$SUPA, School of Physics \& Astronomy, University of St Andrews, North Haugh, St Andrews KY16 9SS, UK} 
\affil{$^{7}$Institut d¡¯Astrophysique de Paris, UMR7095 CNRS--Universit\'e Pierre \& Marie Curie, 98 bis boulevard Arago, F-75014 Paris, France} 
\affil{$^{8}$Las Cumbres Observatory Global Telescope Network, 6740B Cortona Dr, Goleta, CA 93117, USA} 
\affil{$^{9}$School of Physics and Astronomy, Queen Mary University of London, Mile End Road, London E1 4NS, UK} 
\affil{$^{10}$INFN, Sezione di Napoli, Italy} 
\affil{$^{11}$Dipartimento di Fisica ¡°E.R. Caianiello¡±, Universit$\grave{\rm a}$ di Salerno, Via Ponte Don Melillo 84084 Fisciano (SA), Italy} 
\affil{$^{12}$Solar-Terrestrial Environment Laboratory, Nagoya University, Nagoya, 464-8601, Japan} 
\affil{$^{13}$Institute of Information and Mathematical Sciences, Massey University, Private Bag 102-904, North Shore Mail Centre, Auckland, New Zealand} 
\affil{$^{14}$Department of Physics, University of Auckland, Private Bag 92-019, Auckland 1001, New Zealand} 
\affil{$^{15}$School of Chemical and Physical Sciences, Victoria University, Wellington, New Zealand} 
\affil{$^{16}$Okayama Astrophysical Observatory, National Astronomical Observatory of Japan, Asakuchi, Okayama 719-0232, Japan} 
\affil{$^{17}$Nagano National College of Technology, Nagano 381-8550, Japan} 
\affil{$^{18}$Tokyo Metropolitan College of Aeronautics, Tokyo 116-8523, Japan} 
\affil{$^{19}$Mt. John University Observatory, P.O. Box 56, Lake Tekapo 8770, New Zealand} 
\affil{$^{20}$Universidad de Concepci'on, Departamento de Astronomia, Casilla 160-C, Concepci\'on, Chile} 
\affil{$^{21}$Institute of Astronomy, University of Cambridge, Madingley Road, Cambridge CB3 0HA, UK} 
\affil{$^{22}$Instituto Nacional de Pesquisas Espaciais, S$\tilde{\rm a}$o Jos\'e dos Campos, SP, Brazil} 
\affil{$^{23}$Department of Physics, Texas A\&M University, College Station, TX 77843, USA} 
\affil{$^{24}$Institute for Advanced Study, Einstein Drive, Princeton, NJ 08540, USA} 
\affil{$^{25}$School of Physics and Astronomy and Wise Observatory, Tel-Aviv University, Tel-Aviv 69978, Israel} 
\affil{$^{26}$Department of Physics, Ohio State University, 191 W. Woodruff, Columbus, OH 43210, USA} 
\affil{$^{27}$Korea Astronomy and Space Science Institute, Daejeon 305-348, Republic of Korea} 
\affil{$^{28}$Departamento de Astronom$\grave{\rm i}$a y Astrof$\grave{\rm i}$sica, Universidad de Valencia, E-46100 Burjassot, Valencia, Spain} 
\affil{$^{29}$Qatar Foundation, P.O. Box 5825, Doha, Qatar} 
\affil{$^{30}$HE Space Operations, Flughafenallee 24, 28199 Bremen, Germany} 
\affil{$^{31}$Istituto Internazionale per gli Alti Studi Scientifici (IIASS), Vietri Sul Mare (SA), Italy} 
\affil{$^{32}$National Astronomical Observatories/Yunnan Observatory, Key Laboratory for the Structure and Evolution of Celestial Objects, Chinese Academy of Sciences, Kunming 650011, China} 
\affil{$^{33}$Institut d¡¯Astrophysique et de G\'eophysique, All\'e du 6 Ao$\hat{\rm u}$t 17, Sart Tilman, B$\hat{\rm a}$t. B5c, 4000 Li\'ege, Belgium} 
\affil{$^{34}$Stellar Astrophysics Center (SAC), Department of Physics and Astronomy, Aarhus University, Ny Munkegade 120, 8000 $\AA$rhus C, Denmark} 
\affil{$^{35}$Niels Bohr Institute, University of Copenhagen, Juliane Maries vej 30, 2100 Copenhagen, Denmark} 
\affil{$^{36}$Centre for Star and Planet Formation, Geological Museum, ¨ªster Voldgade 5, 1350 Copenhagen, Denmark} 
\affil{$^{37}$Institut for Rumforskning og-teknologi, Danmarks Tekniske Universitet, Juliane Maries Vej 30, 2100 K{\o}benhavn, Denmark}   
\affil{$^{38}$Institut f\"{u}r Astrophysik, Georg-August-Universit\"{a}t, Friedrich-Hund-Platz 1, 37077 G\"{o}ttingen, Germany} 
\affil{$^{39}$Nordic Optical Telescope, Apartado 474, E-38700 Santa Cruz de La Palma, Spain} 
\affil{$^{40}$ESO Headquarters, Karl-Schwarzschild-Str. 2, 85748 Garching bei M\"{u}nchen, Germany} 
\affil{$^{41}$Jodrell Bank Centre for Astrophysics, University of Manchester, Oxford Road, Manchester M13 9PL, UK} 
\affil{$^{42}$Astronomisches Rechen-Institut, Zentrum f\"{u}r Astronomie der Universit\"{a}t Heidelberg (ZAH), M\"{o}nchhofstr. 12-14, 69120 Heidelberg, Germany} 
\affil{$^{43}$Max Planck Institute for Astronomy, K\"{o}nigstuhl 17, 69117 Heidelberg, Germany} 
\affil{$^{44}$Department of Physics, Sharif University of Technology, P.O. Box 11155--9161, Tehran, Iran} 
\affil{$^{45}$Perimeter Institue for Theoretical Physics, 31 Caroline St. N., Waterloo, ON N2L2Y5, Canada} 
\affil{$^{46}$INAF/Istituto di Astrofisica Spaziale e Fisica Cosmica - Bologna, Via Gobetti 101, 40129 Bologna, Italy} 
\affil{$^{47}$Max Planck Institute for Solar System Research, Max-Planck-Str. 2, 37191 Katlenburg-Lindau, Germany} 
\affil{$^{48}$European Southern Observatory (ESO), Alonso de Cordova 3107, Casilla 19001, Santiago 19, Chile} 
\affil{$^{49}$Astrophysics Group, Keele University, Staffordshire ST5 5BG, UK} 
\affil{$^{50}$Department of Physics and Astronomy, University of Canterbury, Private Bag 4800, Christchurch 8020, New Zealand} 
\affil{$^{51}$IRAP, Universit\'e de Toulouse, CNRS, 14 Avenue Edouard Belin, 31400 Toulouse, France} 
\affil{$^{52}$UPMC-CNRS, UMR 7095, Institut d¡¯Astrophysique de Paris, 98bis boulevard Arago, F-75014 Paris, France} 
\affil{$^{53}$School of Mathematics and Physics, University of Tasmania, Private Bag 37, Hobart, TAS 7001, Australia} 
\affil{$^{54}$Department of Physics, University of Rijeka, Omladinska 14, 51000 Rijeka, Croatia} 
\affil{$^{55}$Department of Computing, Technical University of Vienna, Wiedner Hauptstrasse, Vienna, Austria} 
\affil{$^{56}$South African Astronomical Observatory, P.O. Box 9, Observatory 7925, South Africa} 
\affil{$^{57}$Space Telescope Science Institute, 3700 San Martin Drive, Baltimore, MD 21218, USA} 
\affil{$^{58}$European Southern Observatory, Karl-Schwarzschild-Str. 2, 85748 Garching bei M\"unchen, Germany} 
\affil{$^{59}$Astrophysics Research Institute, Liverpool John Moores University, Liverpool CH41 1LD, UK} 
\affil{$^{101}$The $\mu$FUN Collaboration} 
\affil{$^{102}$The OGLE Collaboration} 
\affil{$^{103}$The MOA Collaboration} 
\affil{$^{104}$The MiNDSTEp Consortium} 
\affil{$^{105}$The RoboNet Collaboration} 
\affil{$^{106}$The PLANET Collaboration} 
\affil{$^{111}$Royal Society University Research Fellow} 
\affil{$^{112}$Corresponding author}


\begin{abstract}
Although many models have been proposed, the physical mechanisms responsible 
for the formation of low-mass brown dwarfs are poorly understood. The 
multiplicity properties and minimum mass of the brown-dwarf mass function 
provide critical empirical diagnostics of these mechanisms. We present the 
discovery via gravitational microlensing of two very low-mass, very tight 
binary systems. These binaries have directly and precisely measured total 
system masses of 0.025 $M_\odot$ and 0.034 $M_\odot$, and projected separations 
of 0.31 AU and 0.19 AU, making them the lowest-mass and tightest field 
brown-dwarf binaries known. The discovery of a population of such binaries 
indicates that brown dwarf binaries can robustly form at least down to masses 
of $\sim 0.02\ M_\odot$. Future microlensing surveys will measure a mass-selected 
sample of brown-dwarf binary systems, which can then be directly compared to 
similar samples of stellar binaries.
\end{abstract}



\section{Introduction}

Brown dwarfs (BD) are collapsed objects with masses below the minimum mass 
required to fuse hydrogen of $\sim 0.08\ M_\odot$. Direct imaging surveys have 
found that isolated BD systems to be fairly ubiquitous in the field as well as 
in young clusters (see \citet{luhman12} for a review), with frequencies rivaling 
those of their more massive hydrogen-fusing stellar brethren. However, it is 
unclear whether BDs simply represent the low-mass extension of the initial 
collapsed object mass function (IMF), and thus formed via the same processes 
as stars, or if their formation requires additional physical mechanisms.


The minimum mass of the IMF potentially provides an important discriminant 
between various models of BD formation, with very low mass BD binary systems 
being particularly important in this regard. This is because predictions 
for the multiplicity properties of low-mass BDs -- frequency, mass ratio, 
and separation as a function of total system mass and age -- differ significantly 
depending on the formation scenario. Nowever, the currently available 
observational samples are strongly influenced by detection biases and 
selection effects. For example, although BD surveys in young stellar associations 
allow for detections of low-mass BD systems, these associations are typically 
fairly distant, making it difficult to detect tight BD binaries. Conversely, 
field BD binaries can be resolved to much smaller separations, but low-mass, 
old field BDs are quite faint and thus difficult to detect. As a result, the 
current sample of binary BDs is not only small in number but is also substantially 
incomplete, particularly in the regime of low mass and small separation. A further 
complication is that direct mass measurements are available only for a subset of 
tight field BD binaries. Mass estimates of other systems must rely on comparison 
with models, resulting in substantial systematic uncertainties.

Gravitational microlensing is well-suited to fill the gap. Microlensing is the 
astronomical phenomenon wherein the brightness of a star is magnified by the 
bending of light caused by the gravity of an intervening object (lens) located 
between the background star (source) and an observer. Since this effect occurs 
regardless of the lens brightness, microlensing is suitable to detect faint objects 
such as BDs \citep{paczynski86}. For a lensing event produced by a binary lens with 
well resolved brightness variation of the lensed star, it is possible to precisely 
measure the physical parameters of the lensing object including the mass and distance. 
Here we report the discovery and characterization of two binary BD systems, which 
both have very low mass and tight separation, thus constituting a new population.

\section{Observation}

These BD binaries were discovered in microlensing events 
OGLE-2009-BLG-151/MOA-2009-BLG-232 and OGLE-2011-BLG-0420. The events occurred 
on stars located in the Galactic Bulge field with equatorial and Galactic coordinates 
$({\rm RA},{\rm DEC})_{2000}=(17^{\rm h} 54^{\rm m} 22.34^{\rm s}, -29^\circ 03' 20.8'')$, 
$(l,b)_{2000}=(0.88^\circ,-1.70^\circ)$
and 
$({\rm RA},{\rm DEC})_{2000}=(17^{\rm h} 50^{\rm m} 56.18^{\rm s}, -29^\circ 49' 30.2'')$, 
$(l,b)_{2000}=(359.84^\circ,-1.45^\circ)$, respectively.

OGLE-2009-BLG-151/MOA-2009-BLG-232 was first discovered by the Optical Gravitational 
Lensing Experiment (OGLE: \citet{udalski03}) group and was independently discovered 
by the Microlensing Observations in Astrophysics (MOA: \citet{bond01, sumi03}) 
group in 2009 observation season. OGLE-2011-BLG-0420 was detected by the OGLE group in 
the 2011 season. Both events were additionally observed by follow-up observation groups 
including Microlensing Follow-Up Network ($\mu$FUN: \citet{gould06}), Probing Lensing 
Anomalies NETwork (PLANET: \citep{beaulieu06}), RoboNet (\citet{tsapras09}), and 
Microlensing Network for the Detection of Small Terrestrial Exoplanets (MiNDSTEp: 
\citet{dominik10}). In Table \ref{table:one}, we list the survey and follow-up groups 
along with their telescope characteristics. Data reductions were carried out using 
photometry codes developed by the individual groups.

\begin{figure}[ht]
\epsscale{1.15}
\plotone{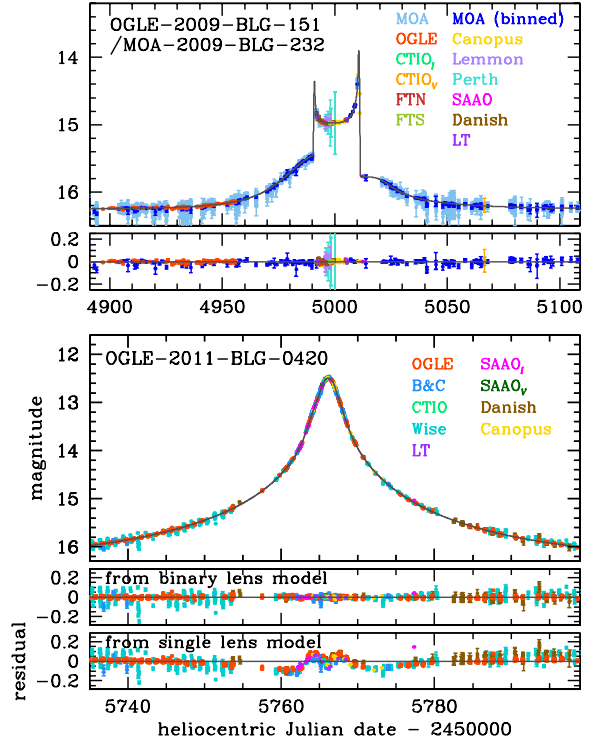}
\caption{\label{fig:one}
Light curves of the binary BD microlensing events
OGLE-2009-BLG-151/MOA-2009-BLG-232 and OGLE-2011-BLG-0420. For the MOA
data of OGLE-2009-BLG-151/MOA-2009-BLG-232, binned data are additionally
plotted to better show residuals.
}\end{figure}

\begin{figure}[ht]
\epsscale{1.15}
\plotone{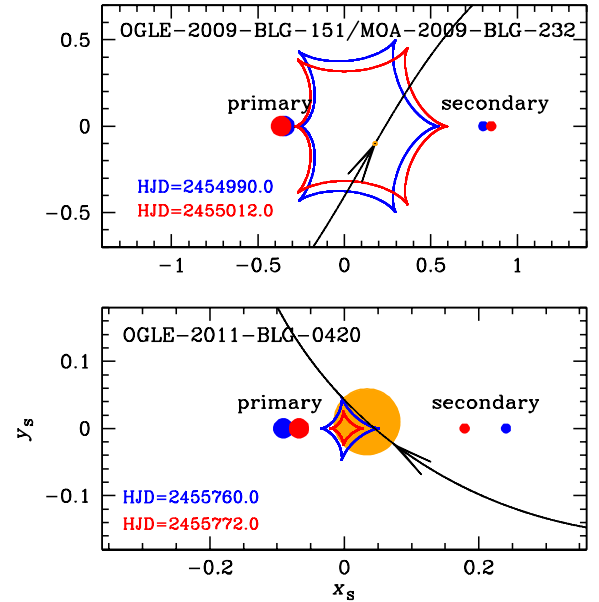}
\caption{\label{fig:two}
Geometry of the lens systems. In each panel, the cuspy closed
figure represents the caustic, the small filled dots are the locations
of the binary lens components, and the curve with an arrow represents
the source trajectory. Two sets of lens positions and the corresponding
caustics are presented at the times marked in the panel. The circle on
the source trajectory represents the scale of the source star with
respect to the caustic. All lengths are normalized by the Einstein
radius corresponding to the total mass of the lens.
}\end{figure}

\begin{deluxetable*}{ll}
\tablecaption{Telescopes\label{table:one}}
\tablewidth{0pt}
\tablehead{
\multicolumn{1}{c}{event} &
\multicolumn{1}{c}{telescope} 
}
\startdata
OGLE-2009-BLG-151  & MOA, 1.8m Mt. John, New Zealand                                   \\
/MOA-2009-BLG-232  & OGLE, 1.3m Warsaw, Las Campanas, Chile                            \\
                   & $\mu$FUN, 1.3m SMARTS, Cerro Tololo Inter-American (CTIO), Chile  \\ 
                   & $\mu$FUN, 1.0m Mt. Lemmon, USA                                    \\
                   & PLANET, 1.0m Canopus, Australia                                   \\
                   & PLANET, 1.0m South African Astronomical (SAAO), South Africa      \\
                   & PLANET, 0.6m Perth, Australia                                     \\
                   & MiNDSTEp, 1.54m Danish, La Silla, Chile                           \\
                   & RoboNet, 2.0m Faulkes North Telescope (FTN), Hawaii, USA          \\
                   & RoboNet, 2.0m Faulkes South Telescope (FTS), Hawaii, USA          \\
                   & RoboNet, 2.0m Liverpool Telescope (LT), Canary Islands, Spain     \\
OGLE-2011-BLG-0420 & OGLE, 1.3m Warsaw, Las Campanas, Chile                            \\
                   & MOA, 0.6m B\&C, Mt. John, New Zealand                             \\
                   & $\mu$FUN, 1.3m SMARTS, CTIO, Chile                                \\
                   & $\mu$FUN, 1.0m Wise, Israel                                       \\
                   & PLANET, 1.0m SAAO, South Africa                                   \\
                   & PLANET, 1.0m Canopus, Australia                                   \\
                   & MiNDSTEp, 1.54m Danish, La Silla, Chile                           \\
                   & RoboNet, 2.0m LT, Canary Islands, Spain
\enddata
\end{deluxetable*}

\section{Analysis}

Figure 1 displays the light curves of the individual events. 
OGLE-2009-BLG-151/MOA-2009-BLG-232 is characterized by two strong spikes flanking 
a ``U''-shape trough, which is typical for caustic-crossing binary-lens events. 
Caustics denote positions on the source plane where the lensing magnification of 
a point source diverges \citep{petters01}. When a caustic is formed by an astronomical object 
composed of 2 masses, it forms a single or multiple sets of closed curves each 
consisting of concave curves that meet at cusps. When a source star crosses the 
caustic, its brightness is greatly enhanced, causing strong deviation from the 
smooth and symmetric single-lens light curve. The light curve of OGLE-2011-BLG-0420, 
on the other hand, appears to be smooth and symmetric, which are the characteristics 
of a lensing event caused by a single mass. From the fit based on the single lens model, 
however, the light curve exhibits noticeable deviations near the peak, which indicates 
the existence of a companion to the lens. According to the classification scheme of 
binary signatures in lensing light curves set by \citet{ingrosso09}, the deviations 
of OGLE-2009-BLG-151/MOA-2009-BLG-232 and OGLE-2011-BLG-0420 are classified as Class II 
and I, respectively.

With known binary signatures, we conduct binary-lens modeling of the observed light curves. 
For a single lens, the light curve is described by 3 parameters: the time of closest lens-source 
approach, $t_0$, the lens-source separation (normalized by the Einstein-ring radius, 
$\theta_{\rm E}$) at that time, $u_0$, and the Einstein time scale, $t_{\rm E}$, 
which represents the time required for the source to cross $\theta_{\rm E}$. The 
Einstein ring denotes the image of a source in the event of perfect lens-source 
alignment, and so is used as the length scale of lensing phenomena. Binary lenses 
require three additional parameters: the mass ratio, $q$, the projected separation 
(normalized by $\theta_{\rm E}$) between the binary components, $s$, and the angle 
between the source trajectory and the binary axis, $\alpha$ (source trajectory angle).

In addition to the basic lensing parameters, it is often needed to include additional 
parameters to precisely describe subtle light-curve features caused by various 
second-order effects. For both events, the lensing-induced magnification lasted for 
several months, which comprises a significant fraction of the Earth's orbital period 
around the Sun (1 year). Then, the apparent motion of the source with respect to the 
lens deviates from rectilinear \citep{gould92} due to the change of the observer's 
position caused by the Earth's orbital motion. This parallax effect causes long-term 
deviation in lensing light curves. Consideration of the parallax effect requires 2 
additional parameters of $\pi_{{\rm E},N}$ and $\pi_{{\rm E},E}$, which are the two 
components of the lens parallax vector $\pivec_{\rm E}$, projected on the sky along 
the north and east equatorial coordinates, respectively.  The orbital motion of the 
lens also affects lensing light curves. The lens orbital motion causes the projected 
binary separation and the source trajectory angle to change over the course of a 
lensing event. These require two additional lensing parameters of the change rates 
of the binary separation, $ds/dt$, and the source trajectory angle, $d\alpha/dt$. 
Finally, finite-source effects become important whenever the magnification varies 
very rapidly with the change of the source position, so that different parts of the 
source are magnified by different amounts. Such a rapid magnification variation occurs 
near caustics and thus finite-source effects are important for binary-lens events 
involved with caustic crossings or approaches. This requires one more parameter, the 
normalized source radius $\rho_*=\theta_*/\theta_{\rm E}$, where $\theta_*$ is the 
angular source radius. Measuring the deviation caused by the parallax and finite-source 
effects is important to determine the physical parameters of the lens.  By measuring 
the finite-source effect, the Einstein radius is determined by $\theta_{\rm E}=
\theta_*/\rho_*$ once the source radius is known. With the measured lens parallax 
and the Einstein radius, the mass and distance to the lens are determined as 
$M_{\rm tot}=\theta_{\rm E}/(\kappa \pi_{\rm E})$ and 
$D_{\rm L}={\rm AU}/(\pi_{\rm E}\theta_{\rm E}+\pi_{\rm S})$,
respectively \citep{gould92, gould06}. Here $\kappa=4G/(c^2{\rm AU})$, AU is an 
Astronomical Unit, $\pi_{\rm S}={\rm AU}/D_{\rm S}$, and $D_{\rm S}\sim 8$ kilo-parsec 
is the source distance.

We model the observed light curves by minimizing $\chi^2$ in the parameter space.
We investigate the existence of possible degenerate solutions because it is known 
that light curves resulting from different combinations of lensing parameters often 
result in a similar shape \citep{griest98, dominik99, an05}. In modeling finite-source 
effects, we additionally consider the limb-darkening variation of the source star 
surface \citep{witt95} by modeling the surface profile as a standard linear law. 
For $\chi^2$ minimization, we use the Markov Chain Monte Carlo method. Photometric 
errors of the individual data sets are rescaled so that $\chi^2$ per degree of 
freedom becomes unity for each data set. We eliminate data points with large errors 
and those lying beyond $3\sigma$ from the best-fit model to minimize their effect 
on modeling.

\begin{deluxetable}{lrr}
\tablecaption{Best fit lensing parameters\label{table:two}}
\tablewidth{0pt}
\tablehead{
\multicolumn{1}{c}{parameters} &
\multicolumn{1}{c}{OGLE-2009-BLG-151} &
\multicolumn{1}{c}{OGLE-2011-BLG-0420} \\
\multicolumn{1}{c}{} &
\multicolumn{1}{c}{/MOA-2009-BLG-232} &
\multicolumn{1}{c}{} 
}
\startdata
$\chi^2$/dof                   &  3040.9/3032           &  5410.3/5439             \\
$t_0$ (HJD')                   &  4999.680 $\pm$ 0.061  &  5766.110 $\pm$ 0.001    \\
$u_0$                          & -0.217 $\pm$ 0.004     & -0.030 $\pm$ 0.001       \\
$t_{\rm E}$ (days)             &  27.95 $\pm$ 0.11      &  35.22 $\pm$ 0.08        \\
$s$                            &  1.135 $\pm$ 0.004     &  0.289 $\pm$ 0.002       \\
$q$                            &  0.419 $\pm$ 0.006     &  0.377 $\pm$ 0.009       \\
$\alpha$                       & -1.049 $\pm$ 0.004     & -2.383 $\pm$ 0.002       \\
$\rho_*$ ($10^{-2}$)           &  1.06 $\pm$ 0.01       &  4.88 $\pm$ 0.01         \\
$\pi_{{\rm E},N}$              & -3.33 $\pm$ 0.11       & -1.15 $\pm$ 0.05         \\
$\pi_{{\rm E},E}$              &  -0.91 $\pm$ 0.11      &  0.19 $\pm$ 0.01         \\
$ds/dt$ (${\rm yr}^{-1}$)      &  1.55 $\pm$ 0.13       & -2.59 $\pm$ 0.07         \\
$d\alpha/dt$ (${\rm yr}^{-1}$) &  0.69 $\pm$ 0.06       &  6.88 $\pm$ 0.20         \\
slope (mag ${\rm yr}^{-1}$)    &  0.0052$\pm$0.0008     &  --   
\enddata
\tablecomments{
HJD'=HJD-2450000.
}
\end{deluxetable}

Table \ref{table:two} gives the solutions of the lensing parameters found from 
modeling. In Figure 2, we also present the geometry of the lens system where the 
source trajectory with respect to the positions of the binary lens components and 
the caustic are shown. For OGLE-2009-BLG-151/MOA-2009-BLG-232, we find that the 
two strong spikes were produced by the source crossings of a big caustic formed 
by a binary lens with the projected separation between the lens components 
($s \sim 1.14$) being similar to the Einstein radius of the lens.  We find that 
including the second-order effects of lens parallax and orbital motions improves 
the fit by $\Delta\chi^2=213$. OGLE-2011-BLG-0420 is also a caustic-crossing event, 
but the projected binary separation ($s\sim 0.29$) is substantially smaller than 
the Einstein radius.  For such a close binary lens, the caustic is small. For 
OGLE-2011-BLG-0420, the caustic is so small that the source size is similar to 
that of the caustic. Hence, the lensing magnification is greatly attenuated by 
the severe finite-source effect and thus the deviation during the caustic crossings 
is weak. We find that there exists an alternative solution with $s > 1$ caused by 
the well-known close/wide binary degeneracy, but the degeneracy is resolved with 
$\Delta\chi^2=27$. The parallax and lens orbital effects are also clearly measured 
with $\Delta\chi^2=403$. From 10 years of OGLE data, we find that OGLE-2011-BLG-0420S 
(source star) is extremely stable, but OGLE-2009-BLG-151/MOA-2009-BLG-232S exhibits 
irregular $< 1\%$ variations, typically on time scales of a few hundred days. Because 
these can affect the parallax measurement, we restrict the modeling to $t_0 \pm 300$ 
days to minimize the impact of variations while still retaining enough baseline to 
ensure a stable fit. We also include a "slope" parameter for the source flux to 
account for the remaining variability. We find only slight differences in final 
results if we repeat this procedure with longer baselines. Therefore, it is unlikely 
but not impossible that source variability affects the 
OGLE-2009-BLG-151/MOA-2009-BLG-232 parallax measurement. By contrast, the results 
for OGLE-2011-BLG-0420 are very robust.

\begin{deluxetable}{lrr}
\tablecaption{Source Star Properties\label{table:three}}
\tablewidth{0pt}
\tablehead{
\multicolumn{1}{c}{quantity} &
\multicolumn{1}{c}{OGLE-2009-BLG-151} &
\multicolumn{1}{c}{OGLE-2011-BLG-0420} \\
\multicolumn{1}{c}{} &
\multicolumn{1}{c}{/MOA-2009-BLG-232} &
\multicolumn{1}{c}{} 
}
\startdata
$(V-I)_0$                &   1.352            &  1.611           \\
$I_0$                    &  14.490            & 13.091           \\
$\theta_*$ ($\mu$as)     &   7.57 $\pm$ 0.66  & 15.94 $\pm$ 1.38 \\
stellar type             &  K giant           & K giant          \\
$\theta_{\rm E}$ (mas)   &  0.71 $\pm$ 0.01   & 0.33 $\pm$ 0.03
\enddata
\end{deluxetable}

\section{Physical Parameters}

\begin{figure}[ht]
\epsscale{1.15}
\plotone{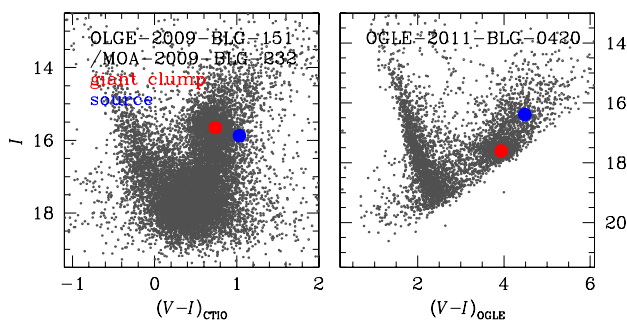}
\caption{\label{fig:three}
Locations of the lensed stars of the individual events on the color-magnitude 
diagrams.
}\end{figure}

Since $\rho_*$ and $\pi_{\rm E}$ are well measured (Table \ref{table:two}), 
it is possible to determine $M_{\rm tot}$ and $D_{\rm L}$ for both systems. 
The only missing ingredient is the angular source radius $\theta_*$, needed 
to find $\theta_{\rm E}=\theta_*/\rho_*$. This is determined from the 
de-reddened color and brightness of the source star. For the calibration of 
the color and brightness, we use the centroid of bulge giant clump as a 
reference \citep{yoo04} because its de-reddened brightness $I_{0,{\rm c}}=14.45$ 
at the Galactocentric distance \citep{nataf12} and the color $(V-I)_{0,{\rm c}}
=1.06$ \citep{bensby11} are known. We then translate $V-I$ into $V-K$ color by 
using the relation \citep{bessell88} and then find $\theta_*$ using the relation 
between the $V-K$ and the angular radius \citep{kervella04}. In Table 
\ref{table:three}, we list the measured de-reddened colors $(V-I)_0$, magnitudes 
$I_0$, angular radii, types of the source stars and the measured Einstein radii 
of the individual events.  Figure \ref{fig:three} shows the locations of the 
lensed stars of the individual events on the color-magnitude diagrams.

\begin{deluxetable}{lrr}
\tablecaption{Physical quantities\label{table:four}}
\tablewidth{0pt}
\tablehead{
\multicolumn{1}{c}{quantity} &
\multicolumn{1}{c}{OGLE-2009-BLG-151} &
\multicolumn{1}{c}{OGLE-2011-BLG-0420} \\
\multicolumn{1}{c}{} &
\multicolumn{1}{c}{/MOA-2009-BLG-232} &
\multicolumn{1}{c}{} 
}
\startdata
$M_{\rm tot}$ ($M_\odot$)      & 0.025 $\pm$ 0.001    & 0.034 $\pm$ 0.002    \\
$M_1$ ($M_\odot$)              & 0.018 $\pm$ 0.001    & 0.025 $\pm$ 0.001    \\
$M_2$ ($M_\odot$)              & 0.0075 $\pm$ 0.0003  & 0.0094 $\pm$ 0.0005  \\
$D_{\rm L}$ (kpc)              & 0.39 $\pm$ 0.01      & 1.99 $\pm$ 0.08      \\
$d_\perp$ (AU)                 & 0.31 $\pm$ 0.01      & 0.19 $\pm$ 0.01
\enddata
\end{deluxetable}

The derived physical quantities for the OGLE-2009-BLG-151/MOA-2009-BLG-232L 
and OGLE-2011-BLG-0420L binaries are listed in Table \ref{table:four}. Here 
the letter "L" at the end of each event indicates the lens of the event. 
The total system masses are $M_{\rm tot}=(0.025 \pm 0.001)\ M_\odot$ and 
$(0.034 \pm 0.002)\ M_\odot$, respectively, well below the hydrogen-burning 
limit. The projected separations and mass ratios are $d_\perp=(0.31\pm0.01)$ 
AU and $q=0.419 \pm 0.006$ for OGLE-2009-BLG-151/MOA-2009-BLG-232L, and 
$d_\perp=(0.19 \pm 0.01)$ AU and $q=0.377 \pm 0.009$ for OGLE-2011-BLG-0420L. 
It is worth emphasizing the high precisions (< 10\%) with which the total 
system masses and individual component masses are determined.

\begin{figure}[t]
\epsscale{1.15}
\plotone{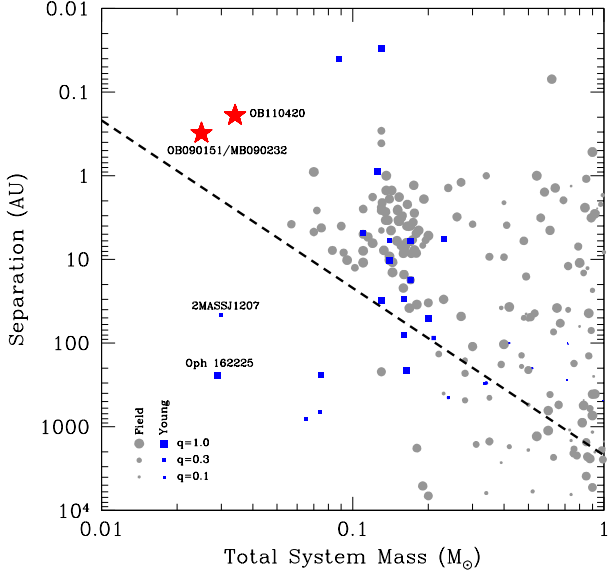}
\caption{\label{fig:four}
Projected separation versus total system mass for a compilation of binaries.  
Grey circles indicate old field binaries, whereas blue squares indicate young 
($< 500$ Myr) systems. The size of the symbols is proportional to the square 
root of the mass ratio. The red stars are the two tight, low-mass binary BDs
discussed here, which have mass ratios of $\sim 0.4$. The dashed line shows a
binding energy of $2 \times 10^{42}$ erg, assuming a mass ratio of 1.
}\end{figure}

Figure \ref{fig:four} compares OGLE-2009-BLG-151/MOA-2009-BLG-232L and 
OGLE-2011-BLG-0420L to a sample of low-mass binaries in the field and in 
young associations from \citet{faherty11}, \citet{basri99}, \citet{burgasser08}, 
\citet{burgasser12}, and \citet{lane01}.  The only known BD binaries with 
comparable total masses are Oph 16225-240515 with $M_{\rm tot}\sim0.032\ 
M_\odot$ \citep{jayawardhana06} and 2MASSJ1207334-393254 \citep{chauvin04} 
with $M_{\rm tot}\sim0.028\ M_\odot$. However, these two systems are both 
young (5 Myr and 8 Myr, respectively), and have very wide separations of 
hundreds of AU. Indeed, OGLE-2009-BLG-151/MOA-2009-BLG-232L and OGLE-2011-BLG-0420L 
are the tightest known BD binaries with substantially lower mass than previously 
known field BD binaries. Both systems have mass ratios of $\sim 0.4$, apparently 
consistent with the trend found from old field BDs, which tend to have a preference 
for larger mass ratios (see Figure \ref{fig:three} of \citet{burgasser07}), 
although it is important to stress that the selection effects in microlensing 
and direct imaging surveys are very different.

\begin{figure}[t]
\epsscale{1.15}
\plotone{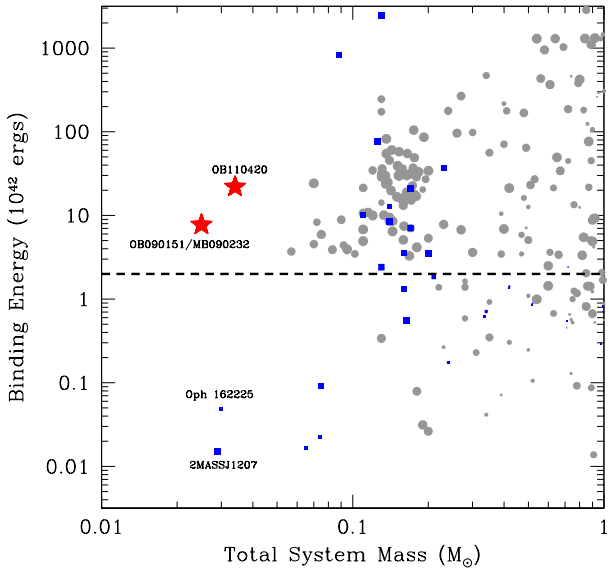}
\caption{\label{fig:five}
Binding energy ($Gm_1m_2/a$) versus total system mass for the same binaries as 
shown in Fig. \ref{fig:four}. Symbols and line are the same.
}\end{figure}

\citet{burgasser07} suggested that field low-mass binaries with $M_{\rm tot}
=0.05$ -- 0.2 $M_\odot$ may exhibit an empirical lower limit to their binding 
energies of $Gm_1m_2/a \sim 2.5\times 10^{42}$ erg (see Figure \ref{fig:five}). 
Although OGLE-2009-BLG-151/MOA-2009-BLG-232L and OGLE-2011-BLG-0420L are 
substantially lower in mass than these BD binaries, they are also considerably 
tighter.  Therefore, with binding energies of $\sim 7\times 10^{42}$ erg and 
$2\times 10^{43}$ erg, they are consistent with the extrapolation of the minimum 
binding energy limit down to total system masses of $M_{\rm tot}\sim 0.02\ M_\odot$.

Although we are unable to provide an estimate of the space density of such tight, 
low-mass brown dwarf binaries, nor even an estimate of their frequency relative 
to more massive stellar binaries, the discovery of two systems among the relatively 
small sample of binary lensing events with precise mass estimates strongly suggests 
that very low-mass, very tight BD binaries are not rare. Thus these detections 
herald a much larger population of such systems. We can therefore conclude that 
BD binaries can robustly form at least down to system masses of $\sim 0.02\ M_\odot$, 
providing a strong constraint for formation models.

\section{Discussion}

The discoveries of the binary BDs reported in this paper demonstrate the importance of 
microlensing in BD studies. The microlensing method has various advantages. First, it 
enables to detect faint old populations of BDs that could not be studied by the conventional
 method of imaging surveys and the sensitivity extends down to planetary mass objects 
\citep{sumi11}.  It also allows one to detect BDs distributed throughout the Galaxy. 
Therefore, microlensing enables to study BDs based on a sample that is not biased by 
the brightness and distance.  Second, in many cases of microlensing BDs, it is possible 
to precisely measure the mass, which is not only the most fundamental physical parameter 
but also a quantity enabling to unambiguously distinguish BDs from other low mass 
populations such as low mass stars.  While mass measurements by the conventional method 
require long-term and multiple stage observation of imaging, astrometry, and spectroscopy 
by using space-borne or very large ground-based telescopes, microlensing requires simple 
photometry by using 1 m class telescopes. Despite the observational simplicity, the mass 
can be measured with uncertainties equivalent to or smaller than those of the measurement 
by conventional methods. Finally, microlensing can expand the ranges of masses and 
separations in the binary BD sample that is incomplete below $\sim 0.1\ M_\odot$ in mass 
and $\sim 3$ AU in separation. Microlensing sensitivity to binary objects peaks when the 
separation is of order the Einstein radius. Considering that the Einstein radius corresponding 
to a typical binary BD is $< 1$ AU, microlensing method will make it possible to study binary 
BDs with small separations.

The number of microlensing BDs is expected to increase in the future with the upgrade 
of instruments in the existing survey experiments and the advent of new surveys. The 
OGLE group recently upgraded its camera with a wider field of view to significantly 
increase the observational cadence. The Korea Microlensing Telescope Network (KMTNet), 
now being constructed, will achieve 10 minute sampling of all lensing events by using 
a network of 1.6 m telescopes on three different continents in the Southern hemisphere 
with wide-field cameras. Furthermore, there are planned lensing surveys in space including 
EUCLID and WFIRST. With the increase of the microlensing event rate combined with the 
improved precision of observation, microlensing will become a major method to study BDs.

\acknowledgments 
Work by C.Han was supported by the research grant of Chungbuk National University in 2011.
The OGLE project has received funding from the European Research Council under 
the European Community¡¯s Seventh Framework Programme (FP7/2007-2013) / ERC grant 
agreement no. 246678. The MOA experiment was supported by grants JSPS22403003 and 
JSPS23340064. This work is based in part on data collected by MiNDSTEp with the Danish 
1.54m telescope at the ESO La Silla Observatory, which is operated based on a grant 
from the Danish Natural Science Foundation (FNU). The MiNDSTEp monitoring campaign 
is powered by ARTEMiS (Dominik et al. 2008, AN 329, 248). MH acknowledges support 
by the German Research Foundation (DFG). DR (boursier FRIA), FF (boursier ARC) and JSurdej 
acknowledge support from the Communaut\'e fran\c{c}aise de Belgique -- Actions de 
recherche concert\'ees -- Acad\'emie universitaire Wallonie-Europe. KA, DMB, MD, 
KH, MH, CL, CS, RAS, and YT are thankful to Qatar National Research Fund (QNRF), 
member of Qatar Foundation, for support by grant NPRP 09-476-1-078. CS has 
received funding from FP7/2007-2013 under grant agreement no. 268421. KH is 
supported by a Royal Society Leverhulme Trust Senior Research Fellowship. 
AG and BSG acknowledge support from NSF AST-1103471. BSG, AG, and RWP acknowledge 
support from NASA grant NNX12AB99G. Work by JCY is supported by a National Science 
Foundation Graduate Research Fellowship under Grant No. 2009068160. SDong¡¯s 
research was performed under contract with the California Institute of Technology 
funded by NASA through the Sagan Fellowship Program. TS was supported by the grant 
JSPS23340044. YM acknowledges support from JSPS grants JSPS23540339 and JSPS19340058. 
TCH and CUL acknowledge the support of KASI grant 2012-1-410-02 and KRCF.

\end{document}